# Chern insulator phase realized in dual-gate-tuned MnBi$_2$Te$_4$ thin films grown by molecular beam epitaxy

Yunhe Bai[1†], Yuanzhao Li[1†], Ruixuan Liu[1†], Jianli Luan[1†], Yang Chen[1], Wenyu Song[1], Peng-Fei Ji[1], Cui Ding[1,3], Zongwei Gao[3], Qinghua Zhang[7], Fanqi Meng[6], Bingbing Tong[3], Lin Li[3], Tianchen Zhu[1], Lin Gu[6], Lili Wang[1,2], Jinsong Zhang[1,2,5], Yayu Wang[1,2,5], Qi-Kun Xue[1,2,3,4], Ke He[1,2,3,5]*, Yang Feng[3]*, and Xiao Feng[1,2,3,5]*

[1]*State Key Laboratory of Low Dimensional Quantum Physics, Department of Physics, Tsinghua University, Beijing 100084, China*
[2]*Frontier Science Center for Quantum Information, Beijing 100084, China*
[3]*Beijing Academy of Quantum Information Sciences, Beijing 100193, China*
[4]*Southern University of Science and Technology, Shenzhen 518055, China*
[5]*Hefei National Laboratory, Hefei 230088, China*
[6] *School of Materials Science and Engineering, Tsinghua University, Beijing 100084, China*
[7]*Institute of Physics, Chinese Academy of Sciences, Beijing 100190, China*

[†] *These authors contributed equally to this work.*

* *Corresponding author. Email:* kehe@tsinghua.edu.cn (K. H.); fengyang@baqis.ac.cn (Y. F.); xiaofeng@mail.tsinghua.edu.cn  (X. F.)

## Abstract

The intrinsic magnetic order, large topological-magnetic gap and rich topological phases make MnBi$_2$Te$_4$ a wonderful platform to study exotic topological quantum states such as axion insulator and Chern insulator. To realize and manipulate these topological phases in a MnBi$_2$Te$_4$ thin film, precise manipulation of the electric field across the film is essential, which requires a dual-gate structure. In this work, we achieve dual-gate tuning of MnBi$_2$Te$_4$ thin films grown with molecular beam epitaxy on SrTiO$_3$(111) substrates by applying the substrate and an AlO$_x$ layer as the gate dielectrics of bottom and top gates, respectively. Under magnetic field of ± 9T and temperature of 20 mK, the Hall and longitudinal resistivities of the films show inversed gate-voltage dependence, for both top- and bottom-gates, signifying the existence of the dissipationless edge state contributed by Chern insulator phase in the ferromagnetic configuration. The maximum of the Hall resistivity only reaches 0.8 $h/e^2$, even with dual-gate tuning, probably due to the high density of bulk carriers introduced by secondary phases. In the antiferromagnetic state under zero magnetic field, the films show normal insulator behavior. The dual-gated

**MnBi$_2$Te$_4$ thin films lay the foundation for developing devices based on electrically tunable topological quantum states.**

The quantum anomalous Hall effect (QAHE), the quantized version of the anomalous Hall effect (AHE), has been proposed and realized in topological insulators with the time-reversal symmetry broken by magnetic order [1-4]. As a quantum Hall effect (QHE) without needing external magnetic field, the chiral edge states contributed by non-trivial bulk topology can support ballistic transport in a long distance, opening an avenue to low-dissipation electronics, quantum computation, and other novel quantum phenomena and applications [1-4]. Compared with magnetically doped topological insulators (TIs)—the first experimentally obtained QAHE system [5], intrinsic magnetic TI MnBi$_2$Te$_4$ is freed from disorder induced by magnetic dopants, providing a better platform for QAHE-related studies [6–14]. MnBi$_2$Te$_4$ has an A-type antiferromagnetic (AFM) ground state with out-of-plane easy magnetic axis (Figure. 1a). Odd- and even-septuple-layer (SL) MnBi$_2$Te$_4$ thin films in the AFM state are QAH and axion insulators, respectively. In the ferromagnetic (FM) state under high magnetic field, MnBi$_2$Te$_4$ becomes a magnetic Weyl semimetal which, in thin films, is supposed to be QAH insulators with Chern number determined by the film thickness [6]. Quantized transport behaviors have been observed in both thin flakes exfoliated from MnBi$_2$Te$_4$ single crystals and molecular beam epitaxy (MBE)-grown MnBi$_2$Te$_4$ thin films [12-22]. With better-controlled sample geometry and size, epitaxial thin film samples are more convenient for systematic studies of the novel quantum effects and easier for construction of MnBi$_2$Te$_4$-based heterostructures for further explorations and applications [21]. But on the other hand, it is much more challenging to obtain quantized transport properties in epitaxial MnBi$_2$Te$_4$ thin films [7, 20-25]. Up to now, quantized Hall resistivity has been only achieved in MnBi$_2$Te$_4$ thin films MBE-grown on sapphire substrates with the chemical potential tuned by controlled oxygen exposure and top-gating [20-22].

Since the quantum effects expected in a MnBi$_2$Te$_4$ thin film are contributed by both its top and bottom surface states, simultaneous tuning of the chemical potential near the two surfaces is highly favored. It is easier to shift the Fermi level into the bulk gap of a MnBi$_2$Te$_4$ thin film by using two gates together, compensating the band-bending effects induced by the substrate and the capping layer. Dual-gate structure can apply an electric field across a magnetic topological insulator thin film, engineering its magnetism and topological phases [26-28]. The electric field generated by dual-gate structure is also essential for realizing and investigating the topological magnetoelectric effect characterizing the axion insulator phase in a MnBi$_2$Te$_4$ thin film [9,18,29]. For MnBi$_2$Te$_4$ thin films grown on sapphire substrates, it is very difficult to make bottom-gate structure. SrTiO$_3$ substrate, with its huge dielectric constant at low temperature (~20000 at 2 K), can be directly used as the gate dielectric of bottom-gate, which has been successfully applied in the magnetically doped $(Bi,Sb)_2Te_3$ thin films. However, the $MnBi_2Te_4$ thin films grown on $SrTiO_3$ substrates reported in existing literatures are all far from the quantized regime [23-25].

In this work, we realize dual-gate-tuned $MnBi_2Te_4$ thin films grown on $SrTiO_3$ substrates which show clear evidences for Chern insulator phase in the FM configuration under high magnetic field. Although the maximal Hall resistivity is still lower than the quantized value (0.8 $h/e^2$), the result demonstrates the system a feasible platform to eventually realize electric field tuning of the QAH and axion insulator states in $MnBi_2Te_4$ thin films.

$MnBi_2Te_4$ thin films are MBE-grown on $SrTiO_3$ (111) substrates with the same growth parameters to the films grown on sapphire [20] (See more details in Experimental Section). Figure 1b displays the reflection high-energy electron diffraction (RHEED) patterns of a typical $SrTiO_3$ substrate (top) and the $MnBi_2Te_4$ thin film grown on it (bottom). The in-plane lattice constant $a$ of $MnBi_2Te_4$ is 4.34 Å, extracted from the streak spacing in RHEED pattern. An atomic force microscopy topographic image of a nominal 5-SL thin film (referred as sample-1 below) capped with cadmium selenide is displayed in Figure 1c, showing a flat terrace with islands and depressions mainly of 1-SL height. X-ray diffraction (XRD) measurement of a 9-SL film shows characteristic diffraction peaks of $MnBi_2Te_4$ phase (Figure 1d) with the out-of-plane lattice constant $c$ about 40.84 Å. Both the in-plane and out-of-plane lattice constants are in good agreement with early works on single crystal samples[11–18]. The energy band structure of a 6-SL $MnBi_2Te_4$ thin film measured by *in-situ* angle-resolved photoemission spectroscopy (ARPES) is similar to that of $MnBi_2Te_4$ bulk crystals, with less electron-doping level (Figure 1e) [11]. *In-situ* oxygen exposure and dual-gate tuning are applied to coarsely and finely tune the Fermi level into the surface state gap, respectively [20]. The dual-gate device configuration and its cross-sectional view are drawn schematically in Figure 1f insets. Figure 1f shows the temperature ($T$)-dependent longitudinal resistivity ($\rho_{xx}$) of sample-1, where a peak appears near 23.4 K, indicating Néel temperature ($T_N$).

Figures 2a and 2b show the magnetic field ($\mu_0H$) dependences of Hall resistivity ($\rho_{yx}$) and longitudinal resistivity ($\rho_{xx}$) of sample-1, respectively, at charge-neutral point (CNP) measured at 20 mK. The CNP is where the top-gate voltage ($V_T$) and bottom-gate voltage ($V_B$) are set such that $|\rho_{yx}|$ and $\rho_{xx}$ reaches the maximum and the minimum, respectively, at ±9 T (see Figures 2c and 2d). At low magnetic field, $\rho_{yx}$-$\mu_0H$ curve exhibits a hysteresis loop, while $\rho_{xx}$-$\mu_0H$ curve shows two peaks near the coercive fields $H_c$. Above 2 T, $|\rho_{yx}|$ grows with increasing magnetic field and saturates above ~7.5 T. At the same time, $\rho_{xx}$ decreases from 1.1 $h/e^2$ at 2 T to 0.54 $h/e^2$ at 7.5 T. The opposite magnetic field dependences of $\rho_{yx}$ and $\rho_{xx}$ suggest the presence of the dissipationless edge state, which means that the film at high magnetic field enters the Chern insulator phase. Nevertheless, even at ±9 T, 20 mK and with both gate-voltages finely tuned, the maximum of $\rho_{yx}$ only reaches 0.8 $h/e^2$, below the quantized value, in contrast to the MnBi$_2$Te$_4$ films grown on sapphire with the same parameters which show fully quantized $\rho_{yx}$ at high magnetic field [20].

To confirm the Chern insulator phase under high magnetic field, we map the dependences of $\rho_{yx}$ and $\rho_{xx}$ on both $V_T$ and $V_B$ at -9 T in Figures 2c and 2d. In the region encircled by the yellow dashed lines, in both Figures 2c and 2d, $\rho_{yx}$ shows maxima, while $\rho_{xx}$ shows minima, in both $V_T$ and $V_B$ dependences. The $\rho_{yx}$ maxima and the $\rho_{xx}$ minima are located at nearly the same ($V_T$, $V_B$) positions (CNPs). The opposite gate-voltage dependences of $\rho_{yx}$ and $\rho_{xx}$ further confirm the existence of the dissipationless edge states of a Chern insulator. It is notable that the CNPs in Figures 2c and 2d roughly construct an oblique line, implying that the top and bottom gates are not independent, but influence each other. It suggests that the electric field applied by the two gates can penetrates the whole film and thus can be used to tune the topological phases. With the guide of Figures 2c and 2d, it is possible to tune the electric field across the film while keeping the overall carrier density nearly constant, as demonstrated in Figures S3c and S3d.

The evolutions of the transport properties related to the Chern insulator phase with temperature and magnetic field are investigated, as shown in Figure 3. The temperature dependences of longitudinal conductivity $\sigma_{xx}$ at different magnetic fields (Figure 3a) all exhibit a thermally activated behavior: $\sigma_{xx} \propto \exp(-\Delta/k_BT)$, as $T$ is lower than ~1.4 K ($k_B$ is Boltzmann constant and $\Delta$ is the size of the activation gap). Figure 3b shows the evolution of the estimated energy gap size $\Delta/k_B$ with magnetic field by Arrhenius fits (solid lines in Figure 3a). $\Delta$ shows two maxima at low field and high field, respectively, with a clear dip between, at ~4.5 T. The observation can be well understood as a transition from normal insulator at AFM state at low field to Chern insulator at FM state under high field, with a delocalization state separating two phases with different Chern numbers [19]. The minimal gap size does not drop to zero, probably due to non-uniform reversal of the magnetic structure. The gap size of the Chern insulator phase at high field is comparable with that of magnetically doped TIs[30], but smaller than that in MnBi$_2$Te$_4$ thin flake samples[12,19]. Figure 3c shows the $\mu_0H$-dependent magnetoresistance curves around 6 T below 0.5 K, where all $\rho_{xx}$ curves cross a fixed point at the critical field $\mu_0H_{cr}$ = 6.17 T, clearly showing a quantum phase transition. The metallic temperature dependence of $\rho_{xx}$ above $H_{cr}$ can be attributed to the edge states of a Chern insulator phase. The inset of Figure 4c shows the scaling analysis of $\rho_{xx}$ versus $(H-H_{cr})/T^\kappa$ near $H_{cr}$, where $\rho_{xx}$ collapse onto a single curve with $\kappa \approx 0.3$. The scaling exponent $\kappa$ is a little different from that in conventional integer QH systems[31] and MnBi$_2$Te$_4$ thin films on sapphire[20], but close to that in magnetically doped TIs with strong disorder[32]. The reason might be related to the parallel dissipative channels or different scattering mechanisms[33], which needs further investigation. The normal insulator behaviors around zero magnetic field were observed in most MnBi$_2$Te$_4$ samples, both thin flakes and thin films, and have been discussed though without a conclusive explanation yet [20].

Finally, we discuss why the MnBi$_2$Te$_4$ thin films grown on SrTiO$_3$ deviate from quantization even under high magnetic field, while the films grown on sapphire under the same conditions can show full quantization. Although the transport properties around high magnetic field and magnetic phase transitions are similar to those observed in MnBi$_2$Te$_4$ thin films on sapphire, the transport data at low magnetic field have notable differences. Figure 4a shows the typical hysteresis loop of $\rho_{yx}$ within ±2 T of a 5-SL MnBi$_2$Te$_4$ thin film on sapphire (sample-3), similar to the samples that show fully quantized $\rho_{yx}$ at 20 mK [20-22]. Compared with the film grown on sapphire, the $\rho_{yx}$ hysteresis loops of the films grown on SrTiO$_3$ (Figure 4b for sample-4 and also Figure 2a for sample-1) always show a kink at a magnetic field below $H_c$ (indicated as $H_c^*$). Correspondingly, in the $\rho_{xx}$-$\mu_0 H$ curve of MnBi$_2$Te$_4$ thin film on SrTiO$_3$ (Figure 4d), a peak is observed also at $H_c^*$, which does not appear in the films grown on sapphire (Figure 3a). These observations imply that there exist other magnetic phases than MnBi$_2$Te$_4$ in the films grown on SrTiO$_3$. The secondary magnetic phases have magnetic coupling with MnBi$_2$Te$_4$, as shown by a series of minor loop measurements which reveal hat $H_c^*$ depends on scanning history (Figure S5g). To investigate the possible origin of $H_c^*$, temperature- and gate-dependent $H_c$ and $H_c^*$ extracted from $\rho_{yx}$–$\mu_0 H$ and $\rho_{xx}$–$\mu_0 H$ curves are shown in Figures 4c and 4d, respectively. $H_c$ and $H_c^*$ show different magnetic ordering temperatures and temperature-dependences, indicating that they are from different magnetic phases. Both $H_c$ and $H_c^*$ have weak gate-voltage dependences. So, the magnetic coupling mechanism of the secondary phases is also not carrier-mediated.

Compared with high-resolution scanning transmission electron microscopy (STEM) images of typical MnBi$_2$Te$_4$ thin films grown on sapphire, quintuple-layer (QL) structures, characteristic of Bi$_2$Te$_3$ phase, are more frequently observed in the films on SrTiO$_3$ (Figure S6). The mixture of SL- and QL- structures, analogous to (MnBi$_2$Te$_4$)$_m$(Bi$_2$Te$_3$)$_n$ superlattices, may show smaller $H_c$ than MnBi$_2$Te$_4$ and contribute to the kink in the $\rho_{yx}$ hysteresis loop [34-36]. Our low-temperature scanning tunneling spectroscopy (STS) measurement on MnBi$_2$Te$_4$ thin films grown on Nb-doped SrTiO$_3$ (111) show an obvious change in the spectra as the tip moves across the step edge of 0.3 nm high, exactly the height difference between a SL and a QL, which is probably the boundary between the different phases (Figure S7).

The secondary phases could have band misalignment with the MnBi$_2$Te$_4$ phase (Figure 4f), leading to extra carriers destroying quantized transport properties. Figure 4e display a summary of the relationship between $\rho_{yx}$ (-8 T) at 1.6 K at CNP and carrier density measured above $T_N$ in different MnBi$_2$Te$_4$ thin film samples (all 5-SL thick), both grown on SrTiO$_3$ (red) and sapphire (blue). The carrier density data are extracted from ordinary Hall effect measurements above $T_N$ [20]. The data points show a clear trend that the samples with lower carrier density have larger $\rho_{yx}$. The MnBi$_2$Te$_4$ films grown on SrTiO$_3$ have obviously higher carrier density and thus smaller $\rho_{yx}$ than the films grown on sapphire, probably resulting from the existence of second phases. Actually, the features of secondary phases have been observed in nearly all MnBi$_2$Te$_4$ films grown on SrTiO$_3$ by different groups. It is still not clear why in MnBi$_2$Te$_4$ thin films MBE-grown on SrTiO$_3$ substrates, it is easy to form second phases. A more detailed study on influence of the surface condition and processing procedure of SrTiO$_3$ substrates on the quality of MnBi$_2$Te$_4$ thin films is highly favored for getting fully quantized MnBi$_2$Te$_4$ thin films with dual-gate tuning.

In conclusion, we have obtained dual-gate-tuned MnBi$_2$Te$_4$ thin films MBE-grown on SrTiO$_3$ substrate which show evidence of the Chern insulator phase at high ±9 T and 20 mK. The maximum of $\rho_{yx}$ can only reach $0.8h/e^2$ even with careful tuning of two gates. We have found that MnBi$_2$Te$_4$ thin films grown on SrTiO$_3$ substrate always show features of secondary phases, which may lead to a lower sample quality than the films grown on sapphire and the deviation from quantization. Improvement of the SrTiO$_3$ surface cleaning procedure is needed to obtain higher-quality dual-gate-tuned MnBi$_2$Te$_4$ thin films that can be used to investigate the topological magnetoelectric effect.

## Experimental Section

**Thin film syntheses and characterizations**

MnBi$_2$Te$_4$ thin films were grown on high-resistance SrTiO$_3$ (111) substrates in an ultrahigh-vacuum MBE system with base pressure better than $2.0\times10^{-10}$ mbar. The substrates were firstly treated in heated de-ionized water for 90 minutes and then transferred into a high-vacuum chamber equipped with a diode laser heater. After degassing at 400 °C for 2 hours, a two-step annealing process was applied, 950 °C annealing for 30 minutes with $P_{O2}$ around $2\times10^{-5}$ mbar followed by 500 °C annealing for 4 hours with $P_{O2}$ around $2\times10^{-4}$ mbar. The treated substrates were transferred ex-situ to deposition chamber and degassed at 400 °C for 30 minutes before growth. High purity Mn (99.9998%), Bi (99.9999%) and Te (99.9999%) were co-evaporated with commercial Knudsen cells. Post-annealing process at growth temperature for 30 minutes was implemented to further improve sample quality. After growth, samples were in-situ exposed to oxygen for an hour. [18] To avoid uncontrolled degradation by air, 8-nm cadmium selenide (CdSe) was capped on top at room temperature. Then the films were patterned into Hall bar with a molybdenum (Mo) hard mask with an effective size of 40 μm × 80 μm. A 40-nm AlO$_x$ layer was deposited on top as a gate dielectric by atomic layer deposition (ALD). 5/20 nm Ti/Au and silver paint were adopted as top and bottom gate electrodes, respectively. The topography was scanned by an atomic force microscope (Bruker, Innova). The c-axis crystalline structure was measured by a commercial high-resolution X-ray diffractometer (Rigaku, SmartLab).

**Transport measurements**

Magneto-transport measurements were performed in a commercial dilution fridge Oxford Instrument Triton 400 with a base temperature below 20 mK. The AC current was applied by the digital source meter Keithley 6221. At the same time the current was measured by the voltage drops on a series 10 kΩ resistor divided by the value of resistance. The longitudinal and Hall voltage drops $V_{xx}$ and $V_{yx}$ were detected simultaneously with AC excitation by using LI5640 lock-in amplifiers made by NF Corporation. The bottom-gate (SrTiO$_3$ dielectric) and top-gate (AlO$_x$ dielectric) voltages were applied by a Keithley 2400 multimeter.

**Acknowledgements**

We thank C. Liu, X. Zhou and Y. Wang for stimulating discussions. We thank Q. Liu for the help of device fabrications. X. Feng would like to acknowledge support by the National Natural Science Foundation of China (Grant No. 92065206 and 12374184). Y. Feng would like to acknowledge support by the NSFC Grant No. 11904053, National Postdoctoral Program for Innovative Talents (Grant No. BX20180079) and China Postdoctoral Science Foundation (Grant No. 2018M641904). J. S. Zhang would like to acknowledge support by National Natural Science Foundation of China (Grant No. 21975140, and No. 51991313), and the Basic Science Center Project of National Natural Science Foundation of China (Grand No. 51788104).

**Conflict with Interest**

The authors declare no conflict with interest.

# References

[1] C.-X. Liu, S.-C. Zhang, X.-L. Qi, *Annu. Rev. Condens. Matter Phys.* **2016**, *7*, 301.

[2] K. He, Y. Wang, Q.-K. Xue, *Annu. Rev. Condens. Matter Phys.* **2016**, *7*, 301.

[3] C.-Z. Chang, Rev. Mod. Phys.

[4]  Y. Tokura, K. Yasuda, A. Tsukazaki, *Nat. Rev. Phys.* **2019**, *1*, 126.

[5]  C.-Z. Chang, J. Zhang, X. Feng, J. Shen, Z. Zhang, M. Guo, K. Li, Y. Ou, P. Wei, L.-L. Wang, Z.-Q. Ji, Y. Feng, S. Ji, X. Chen, J. Jia, X. Dai, Z. Fang, S.-C. Zhang, K. He, Y. Wang, L. Lu, X.-C. Ma, Q.-K. Xue, *Science* **2013**, *340*, 167.

[6]  K. He, npj Quant. Mater. **2020**, *5*, 90.

[7]  Y. Gong, J. Guo, J. Li, K. Zhu, M. Liao, X. Liu, Q. Zhang, L. Gu, L. Tang, X. Feng, D. Zhang, W. Li, C. Song, L. Wang, P. Yu, X. Chen, Y. Wang, H. Yao, W. Duan, Y. Xu, S.-C. Zhang, X. Ma, Q.-K. Xue, K. He, *Chin. Phys. Lett.* **2019**, *36*, 076801.

[8]  J. Li, Y. Li, S. Du, Z. Wang, B.-L. Gu, S.-C. Zhang, K. He, W. Duan, Y. Xu, *Sci. Adv.* **2019**, *5,* eaaw5685.

[9]  D. Zhang, M. Shi, T. Zhu, D. Xing, H. Zhang, J. Wang, *Phys. Rev. Lett.* **2019**, *122*, 206401.

[10] M. M. Otrokov, I. P. Rusinov, M. Blanco-Rey, M. Hoffmann, A. Yu. Vyazovskaya, S. V. Eremeev, A. Ernst, P. M. Echenique, A. Arnau, E. V. Chulkov, *Phys. Rev. Lett.* **2019**, *122*, 107202.

[11] M. M. Otrokov, I. I. Klimovskikh, H. Bentmann, D. Estyunin, A. Zeugner, Z. S. Aliev, S. Gaß, A. U. B. Wolter, A. V. Koroleva, A. M. Shikin, M. Blanco-Rey, M. Hoffmann, I. P. Rusinov, A. Yu. Vyazovskaya, S. V. Eremeev, Yu. M. Koroteev, V. M. Kuznetsov, F. Freyse, J. Sánchez-Barriga, I. R. Amiraslanov, M. B. Babanly, N. T. Mamedov, N. A. Abdullayev, V. N. Zverev, A. Alfonsov, V. Kataev, B. Büchner, E. F. Schwier, S. Kumar, A. Kimura, L. Petaccia, G. Di Santo, R. C. Vidal, S. Schatz, K. Kißner, M. Ünzelmann, C. H. Min, Simon Moser, T. R. F. Peixoto, F. Reinert, A. Ernst, P. M. Echenique, A. Isaeva, E. V. Chulkov, *Nature* **2019**, *576*, 416.

[12] Y. Deng, Y. Yu, M. Z. Shi, Z. Guo, Z. Xu, J. Wang, X. H. Chen, Y. Zhang, *Science* **2020**, *367*, 895

[13] J. Ge, Y. Liu, J. Li, H. Li, T. Luo, Y. Wu, Y. Xu, J. Wang, *Natl. Sci. Rev.* **2020**, *7*, 1280.

[14] C. Liu, Y. Wang, H. Li, Y. Wu, Y. Li, J. Li, K. He, Y. Xu, J. Zhang, Y. Wang, *Nat. Mater.* **2020**, *19*, 522.

[15] Z. Ying, S. Zhang, B. Chen, B. Jia, F. Fei, M. Zhang, H. Zhang, X. Wang, F. Song, *Phys. Rev. B* **2022**, *105*, 085412.

[16] D. Ovchinnikov, X. Huang, Z. Lin, Z. Fei, J. Cai, T. Song, M. He, Q. Jiang, C. Wang, H. Li, Y. Wang, Y. Wu, D. Xiao, J.-H. Chu, J. Yan, C.-Z. Chang, Y.-T. Cui, X. Xu, *Nano Lett.* **2021**, *21*, 2544.

[17] J. Cai, D. Ovchinnikov, Z. Fei, M. He, T. Song, Z. Lin, C. Wang, D. Cobden, J.-H. Chu, Y.-T. Cui, C.-Z. Chang, D. Xiao, J. Yan, X. Xu, *Nat. Commun.* **2022**, *13*, 1668.

[18] A. Gao, Y.-F. Liu, C. Hu, J.-X. Qiu, C. Tzschaschel, B. Ghosh, S.-C. Ho, D. Bérubé, R. Chen, H. Sun, Z. Zhang, X.-Y. Zhang, Y.-X. Wang, N. Wang, Z. Huang, C. Felser, A. Agarwal, T. Ding, H.-J. Tien, A. Akey, J. Gardener, B. Singh, K. Watanabe, T. Taniguchi, K. S. Burch, D. C. Bell, B. B. Zhou, W. Gao, H.-Z. Lu, A. Bansil, H. Lin, T.-R. Chang, L. Fu, Q. Ma, N. Ni, S.-Y. Xu, *Nature* **2021**, *595*, 521.

[19] W. Lin, Y. Feng, Y. Wang, J. Zhu, Z. Lian, H. Zhang, H. Li, Y. Wu, C. Liu, Y. Wang, J. Zhang, Y. Wang, C.-Z. Chen, X. Zhou, J. Shen, *Nat. Commun.* **2022**, *13*, 7714.

[20] Y. Bai, Y. Li, J. Luan, R. Liu, W. Song, Y. Chen, P.-F. Ji, Q. Zhang, F. Meng, B. Tong, L. Li, Y. Jiang, Z. Gao, L. Gu, J. Zhang, Y. Wang, Q.-K. Xue, K. He, Y. Feng, X. Feng, *Natl. Sci. Rev.* **2024**, *11*, nwad189.

[21] Yunhe Bai, Yuanzhao Li, Jianli Luan, Yang Chen, Zongwei Gao, Wenyu Song, Yitian Tong, Jinsong Zhang, Yayu Wang, Junjie Qi, Chui-Zhen Chen, Hua Jiang, X. C. Xie, Ke He, Yang Feng, Xiao Feng, Qi-Kun Xue, (Preprint) arXiv:2404.09083, submitted: Apr **2024**.

[22] Yuanzhao Li, Yunhe Bai, Yang Feng, Jianli Luan, Zongwei Gao, Yang Chen, Yitian Tong, Ruixuan Liu, Su Kong Chong, Kang L. Wang, Xiaodong Zhou, Jian Shen, Jinsong Zhang, Yayu Wang, Chui-Zhen Chen, XinCheng Xie, Xiao Feng, Ke He, Qi-Kun Xue, (Preprint) arXiv:2401.11450, submitted: Jan **2024**.

[23] Y.-F. Zhao, L.-J. Zhou, F. Wang, G. Wang, T. Song, D. Ovchinnikov, H. Yi, R. Mei, K. Wang, M. H. W. Chan, C.-X. Liu, X. Xu, C.-Z. Chang, *Nano Lett.* **2021**, *21*, 7691.

[24] S. Liu, J. Yu, E. Zhang, Z. Li, Q. Sun, Y. Zhang, L. Li, M. Zhao, P. Leng, X. Cao, J. Zou, X. Kou, J. Zang, F. Xiu, (Preprint) arXiv:2110.00540, submitted: Oct **2021**.

[25] P. Chen, Q. Yao, J. Xu, Q. Sun, A. J. Grutter, P. Quarterman, P. P. Balakrishnan, C. J. Kinane, A. J. Caruana, S. Langridge, A. Li, B. Achinuq, E. Heppell, Y. Ji, S. Liu, B. Cui, J. Liu, P. Huang, Z. Liu, G. Yu, F. Xiu, T. Hesjedal, J. Zou, X. Han, H. Zhang, Y. Yang, X. Kou, *Nat. Electron.* **2022**, *6*, 18.

[26] J. Wang, B. Lian, S.-C. Zhang, *Phys. Rev. Lett.* **2015**, *115*, 036805.

[27] S. Du, P. Tang, J. Li, Z. Lin, Y. Xu, W. Duan, A. Rubio, *Phys. Rev. Research* **2020**, *2*, 022025.

[28] C. Lei, A. H. MacDonald, *Phys. Rev. Materials* **2021**, *5*, L051201.

[29] X.-L. Qi, T.L. Hughes, S.-C. Zhang, *Phys. Rev. B* **2008**, *78*, 195424.

[30] A. J. Bestwick, E. J. Fox, X. Kou, L. Pan, K. L. Wang, D. Goldhaber-Gordon, *Phys. Rev. Lett.* **2015**, *114*, 187201.

[31] B. Huckestein, *Rev. Mod. Phys.* **1995**, *67*, 357.

[32] C. Liu, Y. Ou, Y. Feng, G. Jiang, W. Wu, S. Li, Z. Cheng, K. He, X. Ma, Q. Xue, Y. Wang, *Phys. Rev. X* **2020**, *10*, 041063.

[33] S. Koch, R. J. Haug, K. v. Klitzing, K. Ploog, *Phys. Rev. B* **1991**, *43*, 6828.

[34] I. I. Klimovskikh, M. M. Otrokov, D. Estyunin, S. V. Eremeev, S. O. Filnov, A. Koroleva, E. Shevchenko, V. Voroshnin, A. G. Rybkin, I. P. Rusinov, M. Blanco-Rey, M. Hoffmann, Z. S. Aliev, M. B. Babanly, I. R. Amiraslanov, N. A. Abdullayev, V. N. Zverev, A. Kimura, O. E. Tereshchenko, K. A. Kokh, L. Petaccia, G. Di Santo, A. Ernst, P. M. Echenique, N. T. Mamedov, A. M. Shikin, E. V. Chulkov, *npj Quantum Mater.* **2020**, *5*, 54.

[35] J. Wu, F. Liu, M. Sasase, K. Ienaga, Y. Obata, R. Yukawa, K. Horiba, H. Kumigashira, S. Okuma, T. Inoshita, H. Hosono, *Sci. Adv.* **2019**, *5*, eaax9989.

[36] M. Z. Shi, B. Lei, C. S. Zhu, D. H. Ma, J. H. Cui, Z. L. Sun, J. J. Ying, X. H. Chen, *Phys. Rev. B* **2019**, *100*, 155144.

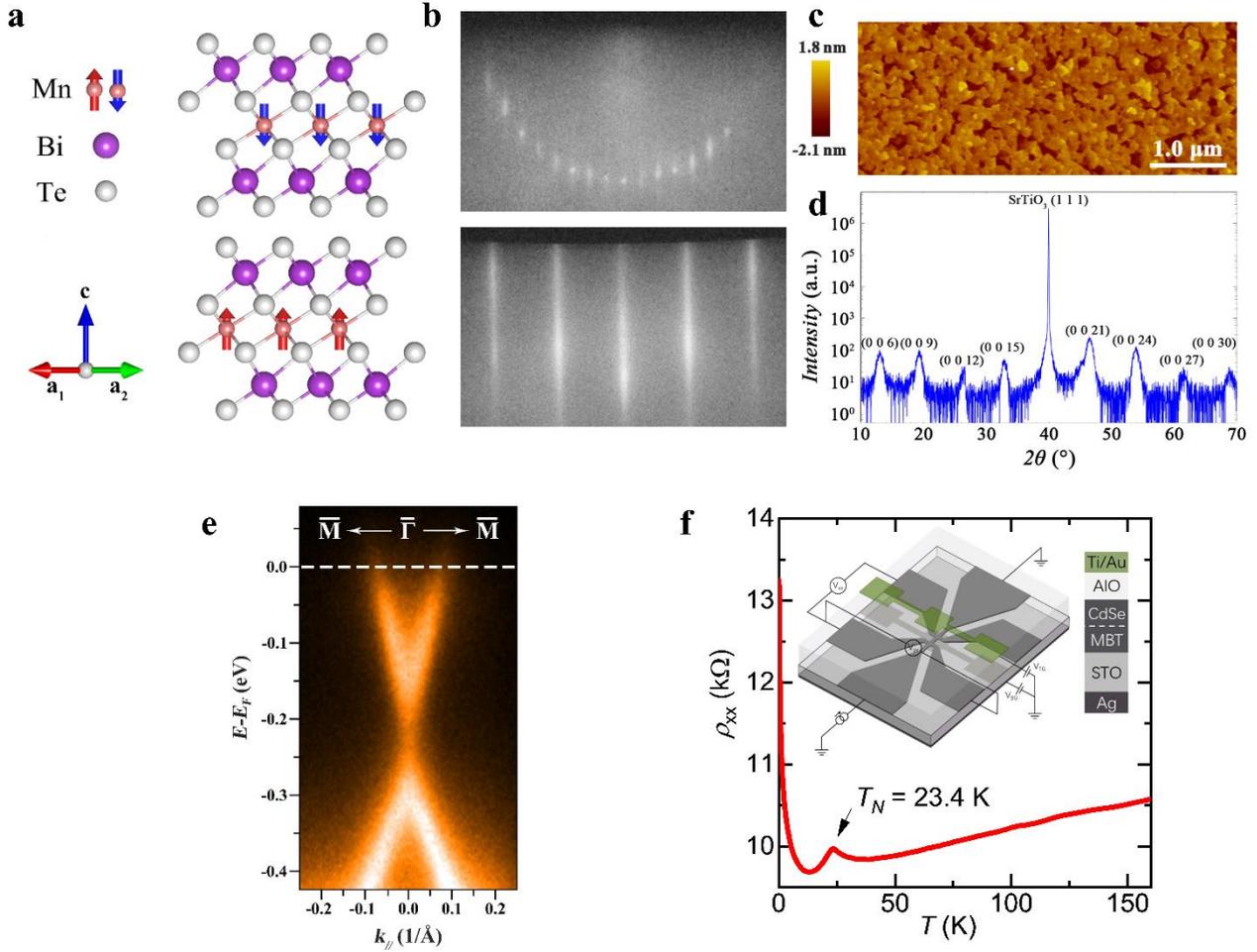

**Figure 1.** Characterizations of epitaxial MnBi$_2$Te$_4$ thin films. a) The schematic crystal structure of tetradymite-type MnBi$_2$Te$_4$. The arrows on manganese atoms denote local moments at zero field. b) RHEED patterns of a typical substrate (top) and a MnBi$_2$Te$_4$ thin film (bottom). c) The topography of the 5-SL MnBi$_2$Te$_4$ film (sample-1) with a capping layer in 2 μm × 5 μm area. d) The XRD pattern of a 9-SL MnBi$_2$Te$_4$ film. e) The electronic band structure of a 6-SL MnBi$_2$Te$_4$ thin film grown on niobium-doped SrTiO$_3$ (111) measured at room temperature. f) The temperature-dependent $\rho_{xx}$ of sample-1 measured down to 100 mK. The AFM phase transition is identified by the peak near $T_N$ = 23.4 K. Insets show the schematic of Hall bar device including the setup of magneto-transport measurements and its cross-section view.

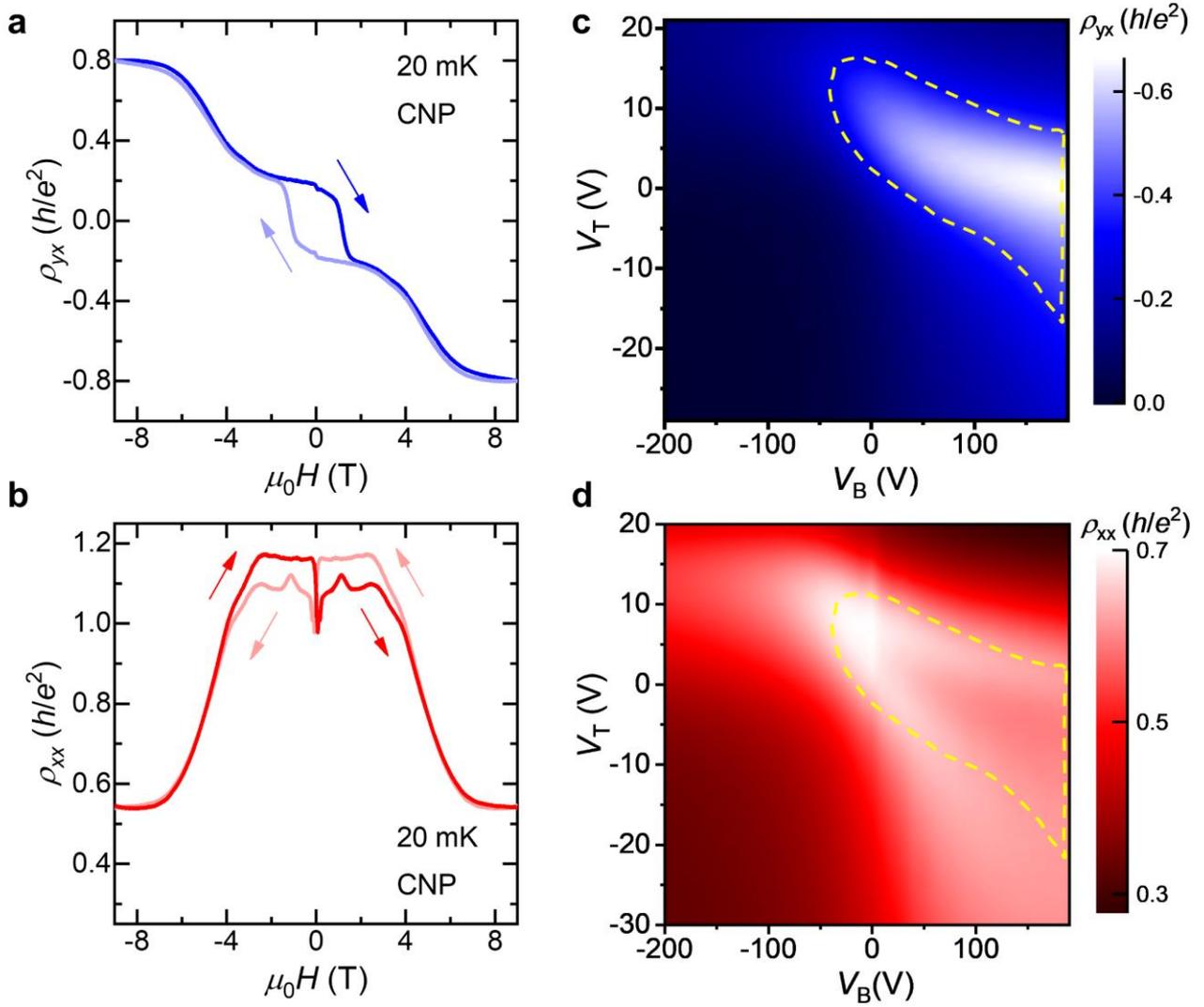

**Figure 2.** Magneto-transport measurements of sample-1 at the base temperature ($T = 20$ mK) with dual-gate tuning. a, b) The magnetic field dependences of $\rho_{yx}$ (a) and $\rho_{xx}$ (b) at CNP, respectively. c, d) Two-dimensional color maps of $\rho_{yx}$ (c) and $\rho_{xx}$ (d) at -9 T, respectively, as a function of $V_B$ and $V_T$. The yellow dashed lines largely denote the regime near Chern insulator phase, corresponding to maxima of $\rho_{yx}$ in (c) and minima of $\rho_{xx}$ in (d).

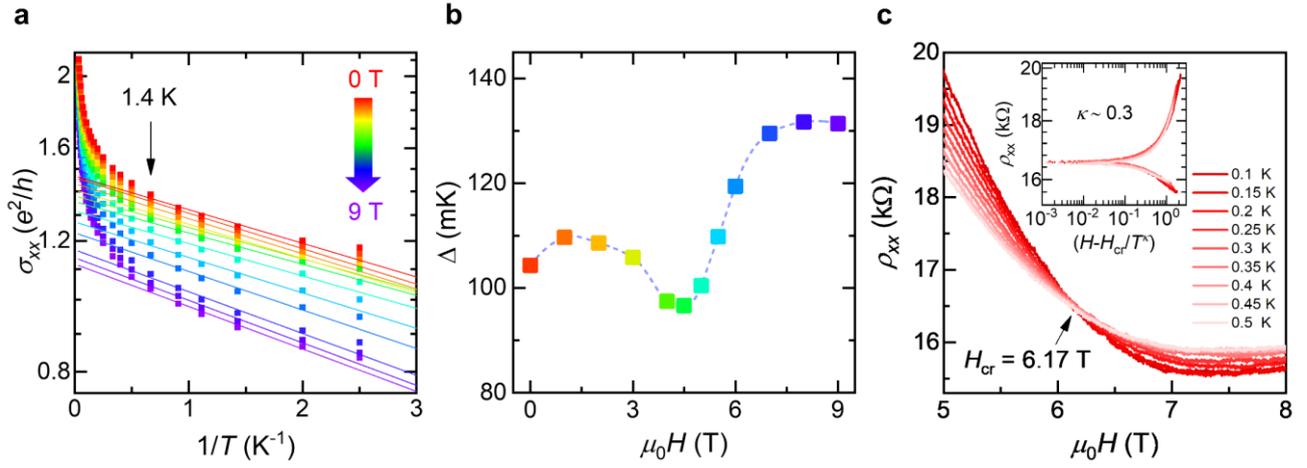

**Figure 3.** The quantum phase transition in sample-1. a) Arrhenius plots of $\sigma_{xx}$ as a function of $1/T$ under different magnetic fields between 0 T and 9 T. Solid lines are linear fits below 1.4 K. b) The activation gap as a function of $\mu_0 H$. The blue dashed line is guide to eye. c) The magnetic field dependence of $\rho_{xx}$ below 0.5 K. The crossing point at $H_{cr} = 6.17$ T is indicated by a black arrow. The inset shows the scaling analysis of $\rho_{xx}$ as a function of $(H-H_{cr})/T^\kappa$.

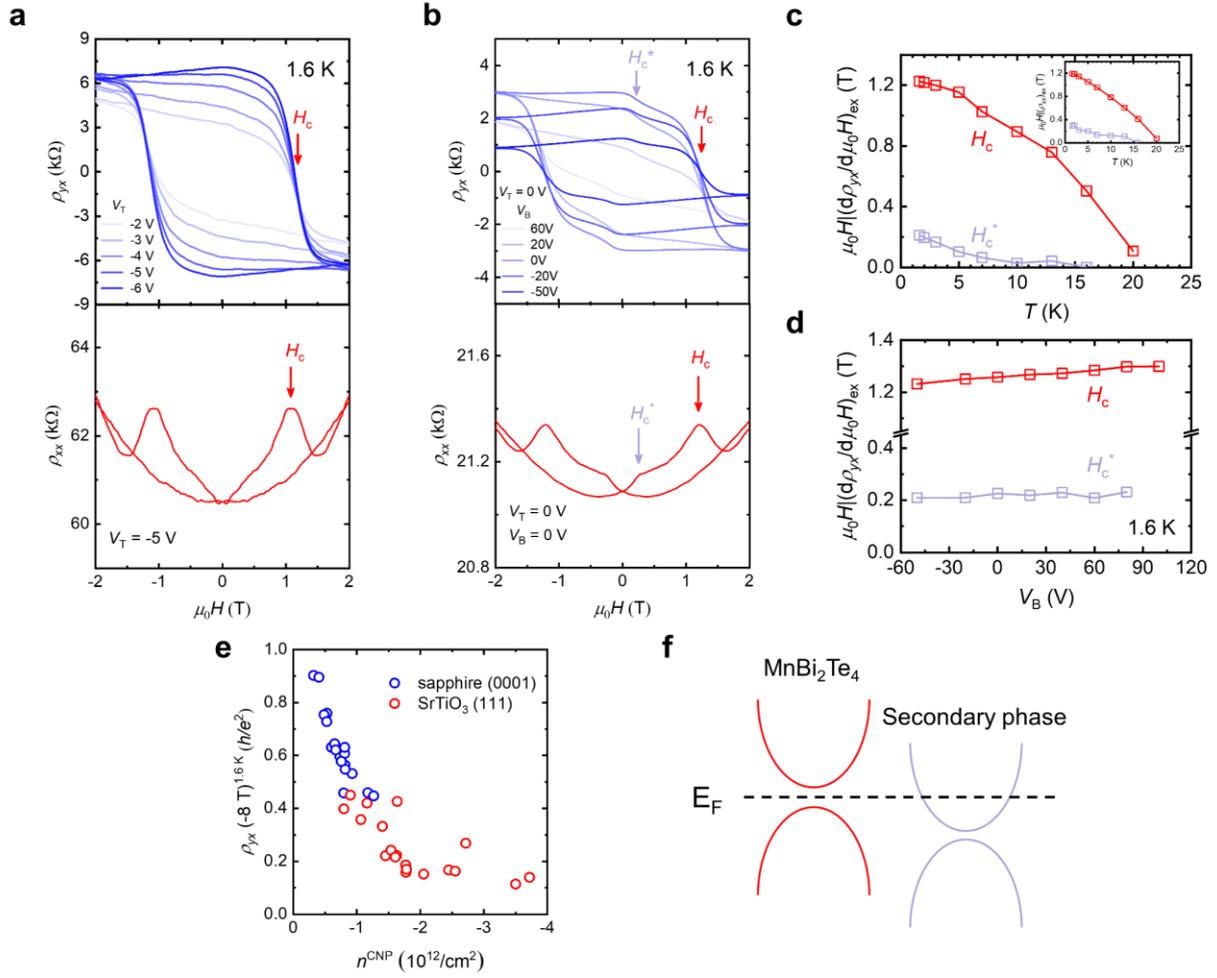

**Figure 4.** The comparison between the magneto-transport behaviors of MnBi$_2$Te$_4$ thin films grown on sapphire and SrTiO$_3$. a), b) Magnetic field-dependent $\rho_{yx}$ and $\rho_{xx}$ of sample-3 (grown on sapphire), sample-4 (grown on SrTiO$_3$) within ±2 T, respectively. c) Temperature-dependent $H_c$ and $H_c^*$ of sample-4 extracted from $\rho_{yx}$–$\mu_0H$ curves. The inset shows the result extracted from $\rho_{xx}$–$\mu_0H$ curves. d) The dependences of $H_c$ and $H_c^*$ on $V_B$ of sample-4 at 1.6 K with $V_T$ =0. e) The statistical result of $\rho_{yx}$ (-8 T) vs. carrier density at CNP of 5-SL MnBi$_2$Te$_4$ thin films grown on sapphire and SrTiO$_3$. The carrier density is extracted from ordinary Hall effect near zero field above $T_N$. f) The schematic of band misalignment of MnBi$_2$Te$_4$ and the secondary phase.

# Supporting Information

# Chern insulator phase realized in dual-gate-tuned MnBi$_2$Te$_4$ thin films grown by molecular beam epitaxy

Yunhe Bai[1†], Yuanzhao Li[1†], Ruixuan Liu[1], Jianli Luan[1†], Yang Chen[1], Wenyu Song[1], Peng-Fei Ji[1], Cui Ding[1,3], Zongwei Gao[3], Qinghua Zhang[7], Fanqi Meng[6], Bingbing Tong[3], Lin Li[3], Tianchen Zhu[1], Lin Gu[6], Lili Wang[1,2], Jinsong Zhang[1,2,5], Yayu Wang[1,2,5], Qi-Kun Xue[1,2,3,4], Ke He[1,2,3,5]*, Yang Feng[3]*, and Xiao Feng[1,2,3,5]*

[1]*State Key Laboratory of Low Dimensional Quantum Physics, Department of Physics, Tsinghua University, Beijing 100084, China*
[2]*Frontier Science Center for Quantum Information, Beijing 100084, China*
[3]*Beijing Academy of Quantum Information Sciences, Beijing 100193, China*
[4]*Southern University of Science and Technology, Shenzhen 518055, China*
[5]*Hefei National Laboratory, Hefei 230088, China*
[6] *School of Materials Science and Engineering, Tsinghua University, Beijing 100084, China*
[7]*Institute of Physics, Chinese Academy of Sciences, Beijing 100190, China*

[†] *These authors contributed equally to this work.*
[*] *Corresponding author. Email:* kehe@tsinghua.edu.cn (K. H.); fengyang@baqis.ac.cn (Y. F.); xiaofeng@mail.tsinghua.edu.cn (X. F.)

1. Magneto-transport results of sample-2
2. Electric field and carrier dependences on $\rho_{yx}$ and $\rho_{xx}$
3. Magnetic phase transitions of sample-1
4. More magneto-transport results of sample-3 and sample-4
5. STEM images of typical thin films grown on SrTiO$_3$ and sapphire
6. STM/STS results of thin films grown on niobium-doped SrTiO$_3$

# 1. Magneto-transport results of sample-2

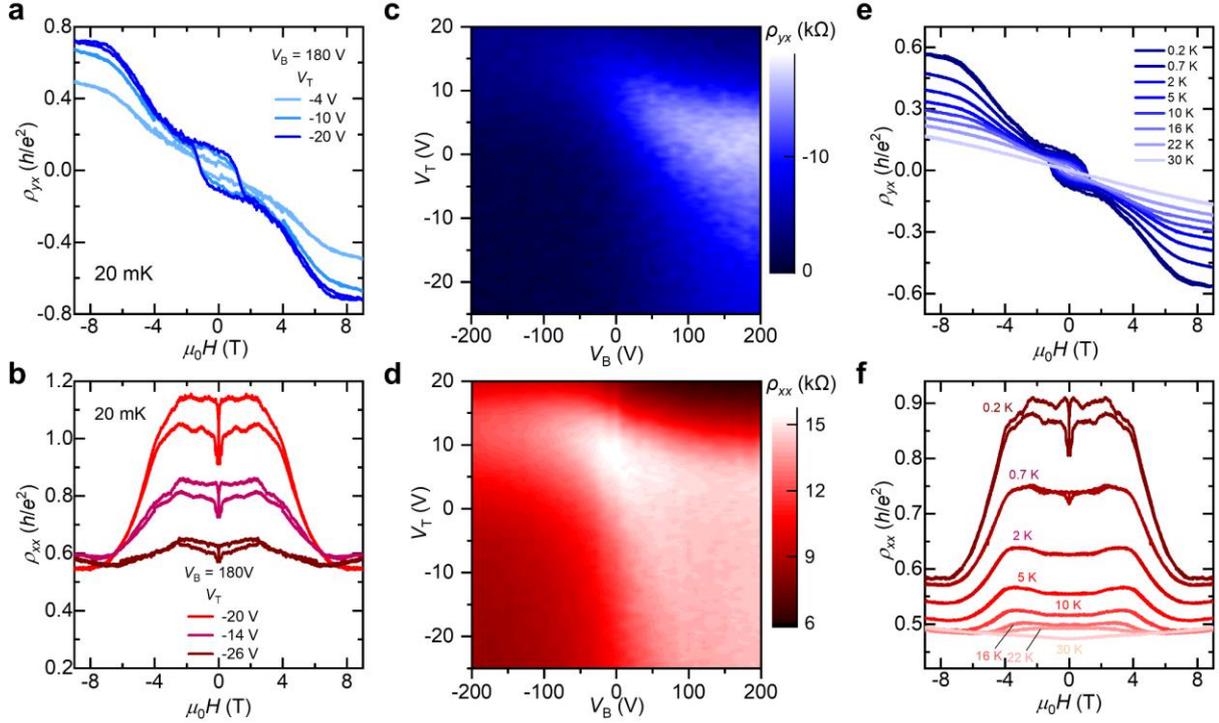

**Figure S1.** Magneto-transport measurements of sample-2. a), b) Magnetic field dependences of $\rho_{yx}$ and $\rho_{xx}$ at 20 mK with selected gate voltages. c), d) Two dimensional color maps of $\rho_{yx}$ (9 T) and $\rho_{xx}$ (9 T) as a function of $V_B$ and $V_T$ at 20 mK. e), f) Temperature dependences of $\rho_{yx}$–$\mu_0 H$ and $\rho_{xx}$–$\mu_0 H$ curves at CNP.

Sample-2, grown with identical growth condition to sample-1 in the main text, basically shows similar transport behaviors. Although the peak of $|\rho_{yx}$ (9 T)$|$ in Figure S1c and the valley of $\rho_{xx}$ (9 T) in Figure S1d is not as obvious as those shown in Figures 2c and 2d, the inverse behavior of $\rho_{yx}$ and $\rho_{xx}$ still exists. This can be clearly seen in Figures S1a and S1b, where an increased $\rho_{yx}$ (-9 T) accompanied by a decreased $\rho_{xx}$ (-9 T) can be observed by tuning $V_T$, indicating a Chern insulator phase. The maximum of $\rho_{yx}$ (-9 T) is about $0.7 h/e^2$ with the minimum of $\rho_{xx}$ (-9 T) about $0.55\ h/e^2$ at $V_B = 180$V, $V_T = -20$V. The temperature evolution of magneto-transport results (Figures S1e and S1f) shows the gradual destruction of magnetic order with increasing temperature.

## 2. Electric field and carrier dependences on $\rho_{yx}$ and $\rho_{xx}$

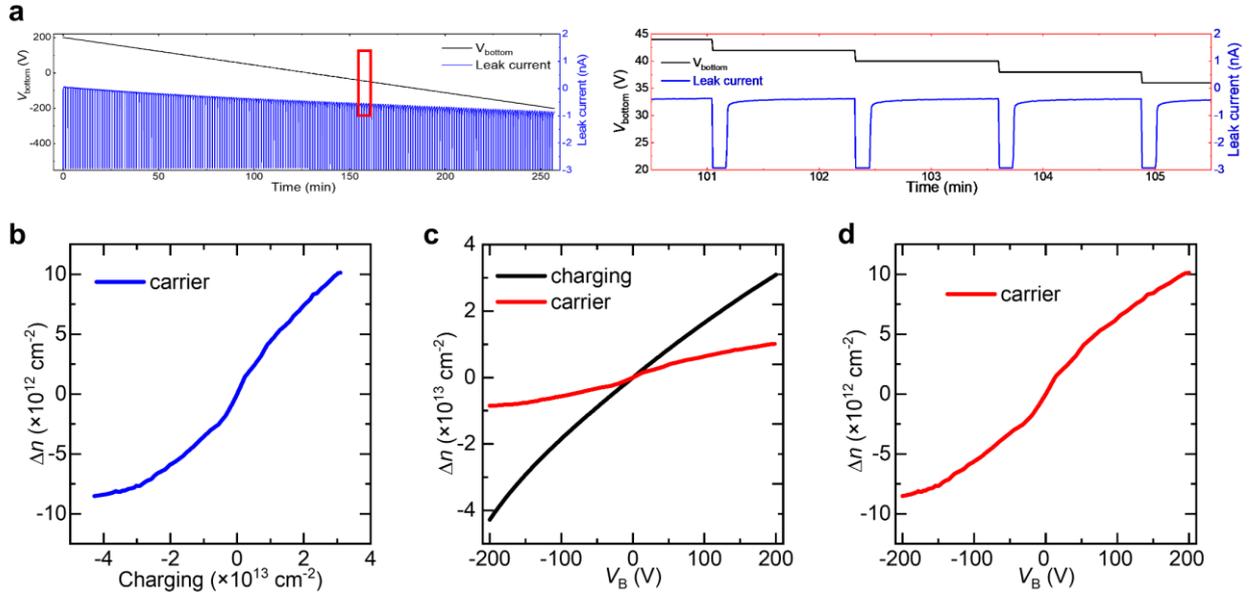

**Figure S2.** Carrier tunability of SrTiO$_3$ (111) substrates. a) $V_B$ and the leak current as a function of time (left). A zoomed-in plot of time-dependent $V_B$ and the leak current in the left red square area (right). b) The relation between the change of carrier density in films and the charge concentration on SrTiO$_3$ dielectric layer extracted from the data of a Bi$_2$Se$_3$ film grown on SrTiO$_3$. c) The changes of charge concentration on SrTiO$_3$ dielectric layer and carrier density in MnBi$_2$Te$_4$ film as a function of $V_B$. The $\Delta n$–$V_B$ curve (red line) is estimated by results in b). d) The estimated change of carrier density in MnBi$_2$Te$_4$ film as a function of $V_B$.

The exact magnitudes of applied vertical electric field and carrier density under different combinations of dual-gate voltages are examined. The carrier density $n = (n_T + n_B) + n_0$ and the electric displacement field $D = (-en_T + en_B)/2\varepsilon_0 + D_0$ can be tuned independently in a dual-gate device, where $n_T$, $n_B$ are carrier densities tuned by top and bottom gates; $\varepsilon_0$ is the vacuum permittivity; $e$ is the electron's charge; $n_0$ and $D_0$ are the initial carrier density and electric displacement field in films. [1] We use $\Delta n$ and $\Delta D$ represent the change of carrier density and displacement field tuned by dual gates in the following. Due to the net magnetization, it is difficult to estimate $\Delta n$ from ordinary Hall effect at low temperature in MnBi$_2$Te$_4$ films directly. $n_T = C_T V_T/eS$ can be easily calculated for an AlO$_x$ dielectric layer, where $C_T$ and $S$ are the capacitance of top dielectric layer and the size of device, respectively. While for $n_B$, it is not easy to determine a reliable value for several reasons. The dielectric constant of SrTiO$_3$ substrate, which is related to $C_B$, is not a constant but a function of $V_B$. [2] $C_B$ will decreases with increasing $V_B$ and $C_B$–$V_B$ curve shows hysteresis between forward and backward sweeps. Moreover, the tuned carriers in thin films and the change of charge concentration $\Delta Q/S$ of the

SrTiO$_3$ dielectric layer are related but not equal, and their relation cannot be described by a simple analytical formula. Thus, to get a reliable dependence of $n_B$ on $V_B$, two steps are needed: 1) determine the relation between $\Delta Q/S$ and $V_B$; 2) determine the relation between $\Delta n_B$ and $\Delta Q/S$.

For the step 1), the leak current and $V_B$ as a function of time were recorded simultaneously as shown in Figure S2a, where $V_B$ was tuned from 200 V to -200 V. The direction of gate sweep is consistent with those in gate-dependent measurements in the main text. The compliance of DC voltage source (Keithley 2400) was set as a small value 3 nA. The value is selected intendedly, since it is much larger than the leak current in the steady state when $V_B$ is fixed, and is smaller than the value of leak current when gate voltage is suddenly changed. The value of leak current switches back and forth between the compliance value (over-leak regime) and a relatively small value (after charging, tails in the right of Figure S2a). The over-leak regime (reach the compliance) is taken into consideration as the effective charging into SrTiO$_3$ capacitor since the main components in the leak current tails arise from the flow going through the insulating coat of wires, i.e. SrTiO$_3$ dielectric layer would be charged by 3 nA for several seconds once $V_B$ is tuned. The relation between $\Delta Q/S$ and $V_B$ is drawn in Figure S2c (the black solid line) by integrated over time. The zero-point of $\Delta Q/S$ is set to the point $V_B = 0$ naturally.

For the step 2), we first study Bi$_2$Se$_3$ films grown on SrTiO$_3$ which shows a linear relationship of $\rho_{yx}$ to $\mu_0 H$ at low temperature. The result is shown in Figure S2b, where $\Delta n$ is extracted from the ordinary Hall effect. Based on the $\Delta n$–$\Delta Q/S$ curve, the estimated $\Delta n$ in MnBi$_2$Te$_4$ thin films can be matched with $V_B$ (Figure S2c), a zoomed-in plot of which is shown in Figure S2d.

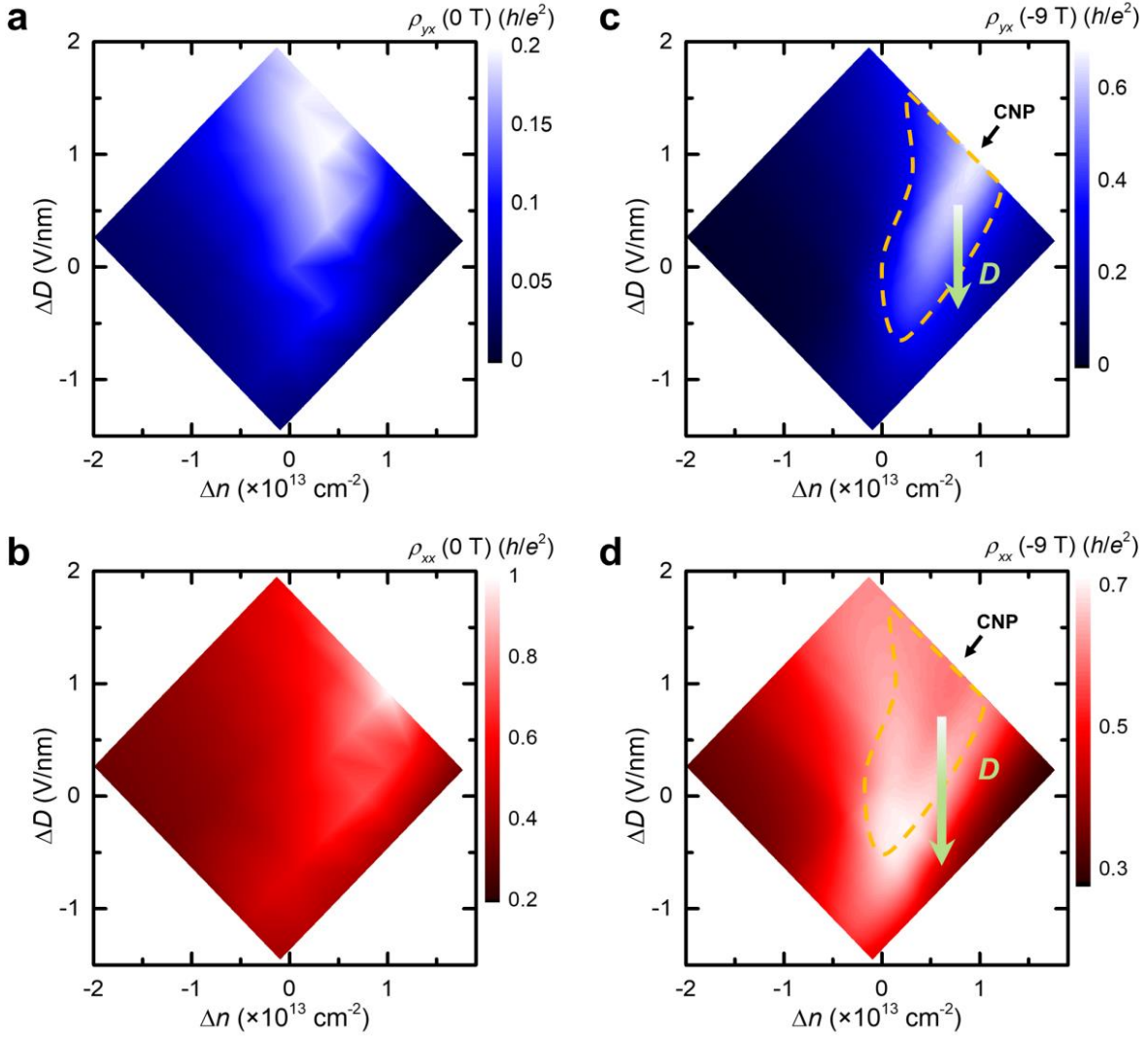

**Figure S3.** Displacement field and carrier density dependences on resistivities of sample-1. a), b), c), d) Two-dimensional color maps of $\rho_{yx}$ (0 T), $\rho_{xx}$ (0 T), $\rho_{yx}$ (-9 T), $\rho_{xx}$ (-9 T), respectively, as a function of $\Delta D$ and $\Delta n$.

With experimentally accessible $n_T$ and $n_B$, $\Delta n$ and $\Delta D$ can be obtained reliably. As shown in Figure S3, both electric filed (along the vertical axis) and carrier density (along the horizontal axis) have an influence on resisitivites. Focus on $\rho_{yx}$ (-9 T) and $\rho_{xx}$ (-9 T) (Figures S3c and S3d), the regions of Chern insulator phase can still be seen largely sketched by yellow dashed line. The CNP indicated by black arrow is located at top right corner rather than the origin point since the axes represent the changes of $n$ and $D$ by dual gates not an absolute value. With a large enough $\Delta D$, electric field can drive the ground state away from the Chern insulator phase, as the green arrows indicated along the vertical axis at a fixed $\Delta n$.

## 3. Magnetic phase transitions of sample-1

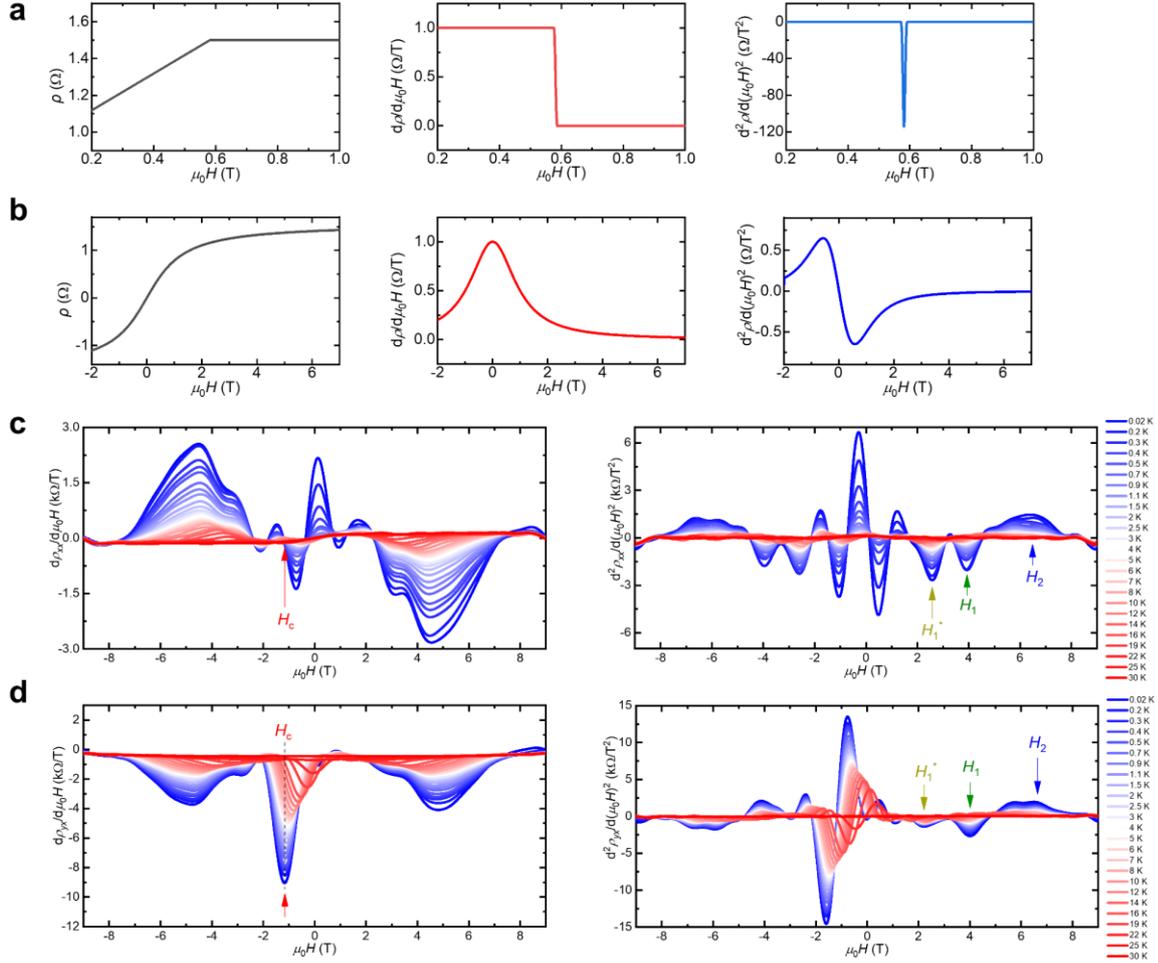

**Figure S4**. Determination of magnetic transitions. a), b) Two examples to elucidate the validy of the data extracting method, a piecewise function in a) and a smooth function in b). c), d) First and second derivatives of $\rho_{xx}$, $\rho_{yx}$ with respect to $\mu_0H$ at various temperatures, respectively. The magnetic phase transition fileds are indicated by arrows with different colors.

The anomalous Hall resistance is related to magnetization in magnetic materials. At CNP, $\rho_{yx}$ keeps a constant when the magnetization is saturated principally. We first introduce a piecewise function $\rho\,(\mu_0H)$ consisted of two straight lines to simulate this behavior:

$$\rho(\mu_0H) = \begin{cases} \mu_0H + 0.923 & \mu_0H < 0.577 \text{ T} \\ 1.5 & \mu_0H \geq 0.577 \text{ T} \end{cases}$$

As shown in the first figure of Figure S4a, $\rho\,(\mu_0H)$ saturates at 0.577 T. The saturation field can be extracted from the first and second derivatives of $\rho\,(\mu_0H)$, showing a step function and a delta function at the same position respectively (Figure S4a). The delta function is smoothed to make the divergence observable. A smoother function arctan($\mu_0H$), which commonly used in the simulations of hysteresis loops,[3] is introduced in Figure S4b to describe a more practical case. The dip in the second derivative of $\rho$ ($\mu_0H$) largely catch the transition field. These two examples elucidate the validity of the data extracting method. The method is also suitable for data extraction from $\rho_{xx}$, since different scattering processes during magnetic phase transitions usually generate an apparent variation in $\rho_{xx}$-$\mu_0H$ curves.

## 4. More magneto-transport results of sample-3 and sample-4

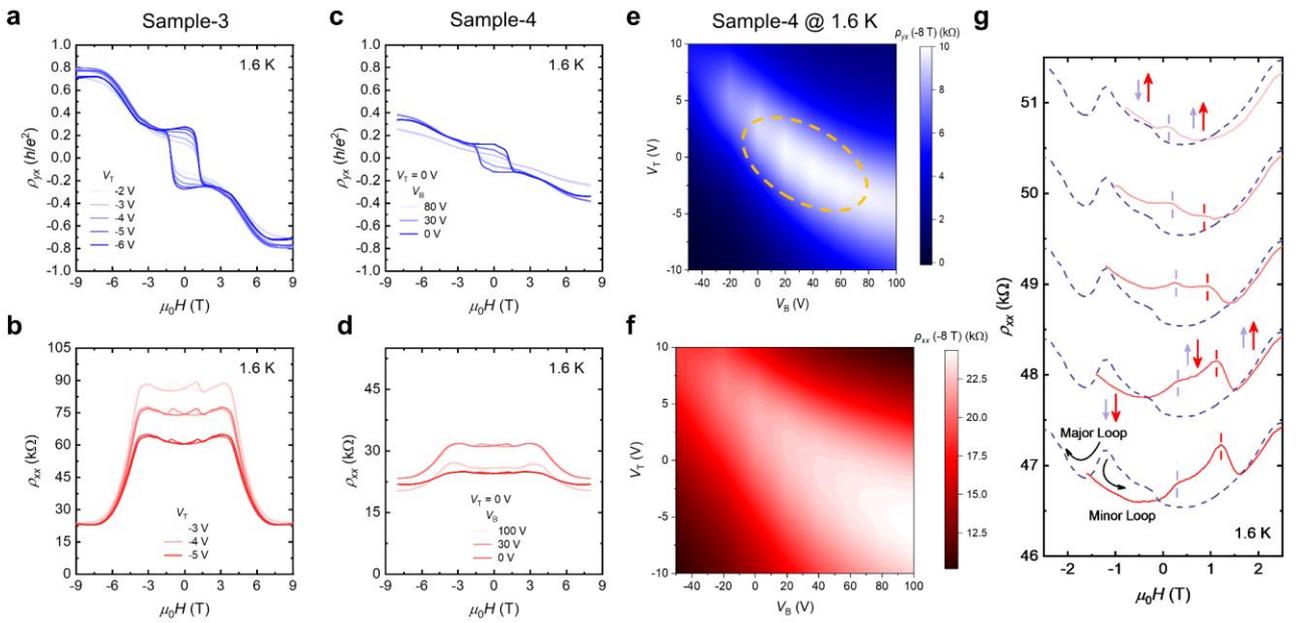

**Figure S5.** Different transport behaviors of sample-3 and sample-4. a), b) Magneto-transport measurements of sample-3 at 1.6 K with different $V_T$s. c), d) Magneto-transport measurements of sample-4 at 1.6 K with different $V_B$s. e), f) Two-dimensional color maps of anti-symmetrized $\rho_{yx}$ (-8 T) and symmetrized $\rho_{xx}$ (-8 T) of sample-4 as a function of $V_B$ and $V_T$, respectively. The yellow dashed line sketches the region of maximum $\rho_{yx}$. g) A series of minor loops of sample-4 at 1.6 K. The blue dashed lines represent the major loop and the red lines represent minor loops with different returning points. The schematic of magnetic configurations of MnBi$_2$Te$_4$ phase (red arrows) and the secondary phase (purple arrows) are drawn in the inset.

# 5. STEM images of typical thin films grown on SrTiO$_3$ and sapphire

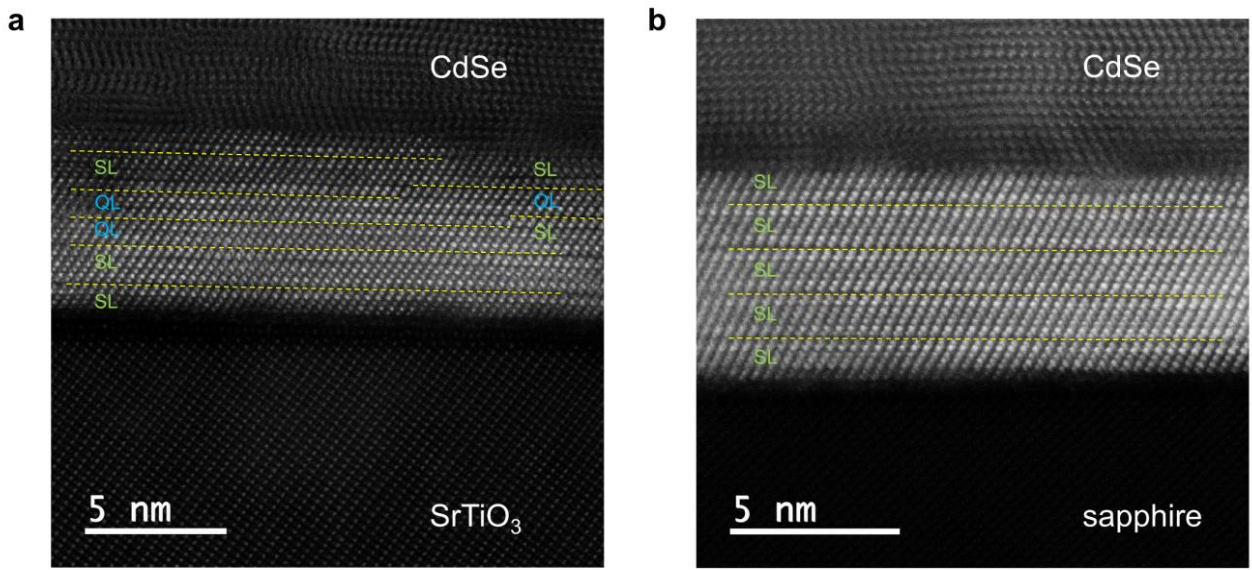

**Figure S6**. Typical STEM images of 5-SL thin films. a), b) Nominal 5-SL MnBi$_2$Te$_4$ thin films grown on SrTiO$_3$ and sapphire, respectively. Yellow dashed lines are guide to eyes.

Figure S6 shows the comparison of typical STEM images of 5-SL thin films grown on SrTiO$_3$ (111) and sapphire (0001). Guide with dashed lines, QL structure can be more easily observed in the former one. Very few QL structure could be found in thin films grown on sapphire, which indicates a much less proportion of the secondary phase. No 5-QL structure had been found in the nominal 5-SL thin films grown on SrTiO$_3$, implying that the formation energy of mixture of SL- and QL-structures is lower than pure Mn-doped Bi$_2$Te$_3$ with a suitable growth parameter.

# 6. STM/STS results of thin films grown on niobium-doped SrTiO$_3$

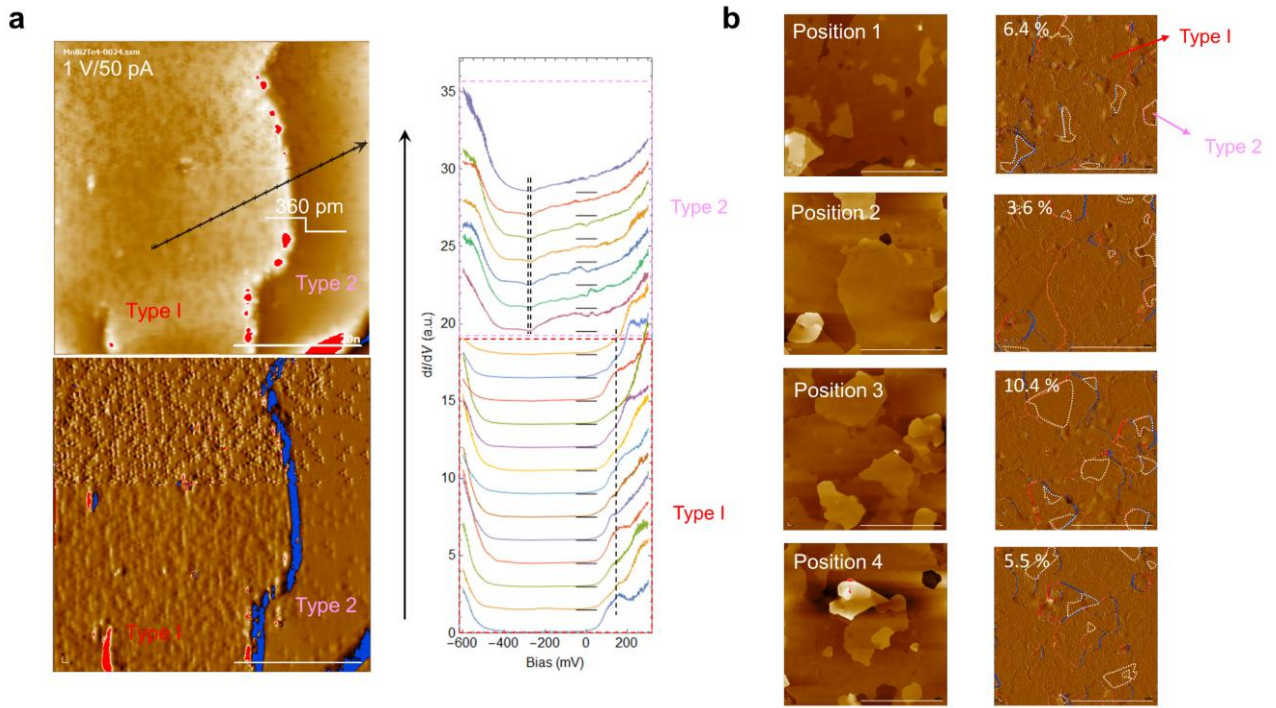

**Figure S7**. STM/STS results of MnBi$_2$Te$_4$ thin films grown on niobium-doped SrTiO$_3$ (111). a). The STM topography in a 50 nm × 50 nm area (top) and its differential image (bottom). Different tunneling spectra scanned along the black line (right). b) Propotions of type-2 phase in different positions. Each scanning area is 400 nm × 400 nm.

In a 50 nm × 50 nm area, two distinct defect concentrations separated by a step, the hieght of which is about 0.36 nm, can be observed in the STM image shown in Figure S7a. The difference can be more obviously seen in the differential image at the bottom. Scanning along the black line, two areas have significantly different tunneling spectra demonstrating their different electronic bands origin from distinguishing phases. Compared with eraly results of STM studies on MnBi$_2$Te$_4$ [4] and Mn-doped Bi$_2$Te$_3$ single crystals,[5] type-1 and type-2 tunneling spectra might be related to MnBi$_2$Te$_4$ and Mn-doped Bi$_2$Te$_3$ phase of topmost layer. According to the difference in defect concentration of two phases, we took a series of STM measurements on different areas of MnBi$_2$Te$_4$ thin films and the portion of type-2 phase is labled at top left corner of each figure in Figure S7b. The STM/STS results indicate the secondary phase could have a significantly different band structure with that of MnBi$_2$Te$_4$.

# References

[1] P. Maher, L. Wang, Y. Gao, C. Forsythe, T. Taniguchi, K. Watanabe, D. Abanin, Z. Papić, P. Cadden-Zimansky, J. Hone, P. Kim, C. R. Dean, *Science* **2014**, *345*, 6192.

[2] H.-M. Christen, J. Mannhart, E. J. Williams, Ch. Gerber, *Phys. Rev. B* **1994**, *49*, 12095.

[3] A. Arrott, *Phys. Rev.* **1957**, *108*, 1394.

[4] Z. Huang, M.-H. Du, J. Yan, W. Wu, *Phys. Rev. Mater.* **2020**, *4*, 121202.

[5] Y. S. Hor, P. Roushan, H. Beidenkopf, J. Seo, D. Qu, J. G. Checkelsky, L. A. Wray, D. Hsieh, Y. Xia, S.-Y. Xu, D. Qian, M. Z. Hasan, N. P. Ong, A. Yazdani, R. J. Cava, *Phys. Rev. B* **2010**, *81*, 195203.